\documentclass[sigconf]{acmart}
\AtBeginDocument{%
  }

\copyrightyear{2026}
\acmYear{2026}
\setcopyright{cc}
\setcctype{by}
\acmConference[CHI '26]{Proceedings of the 2026 CHI Conference on Human Factors in Computing Systems}{April 13--17, 2026}{Barcelona, Spain}
\acmBooktitle{Proceedings of the 2026 CHI Conference on Human Factors in Computing Systems (CHI '26), April 13--17, 2026, Barcelona, Spain}
\acmPrice{}
\acmDOI{10.1145/3772318.3791537}
\acmISBN{979-8-4007-2278-3/2026/04}

\begin{document}

\title{Talking Inspiration: A Discourse Analysis of Data Visualization Podcasts}

\author{Ali Baigelenov}
\orcid{0009-0003-6491-1874}
\affiliation{%
  \department{School of Applied and Creative Computing}
  \institution{Purdue University}
  \city{West Lafayette}
  \state{Indiana}
  \country{USA}
}
\email{abaigele@purdue.edu}

\author{Prakash Shukla}
\orcid{0009-0002-7416-1758}
\affiliation{%
  \department{School of Applied and Creative Computing}
  \institution{Purdue University}
  \city{West Lafayette}
  \state{Indiana}
  \country{USA}
}
\affiliation{%
  \department{School of Architecture \& Design}
  \institution{The University of Kansas}
  \city{Lawrence}
  \state{Kansas}
  \country{USA}}
\email{shukla@ku.edu}

\author{Phuong Bui}
\orcid{0009-0004-3762-4770}
\affiliation{%
  \department{School of Applied and Creative Computing}
  \institution{Purdue University}
  \city{West Lafayette}
  \state{Indiana}
  \country{USA}}
\email{bui42@purdue.edu}

\author{Paul Parsons}
\orcid{0000-0002-4179-9686}
\affiliation{%
  \department{School of Applied and Creative Computing}
  \institution{Purdue University}
  \city{West Lafayette}
  \state{Indiana}
  \country{USA}}
\email{parsonsp@purdue.edu}

\renewcommand{\shortauthors}{Baigelenov et al.}

\begin{abstract}
  
Data visualization practitioners routinely invoke inspiration, yet we know little about how it is constructed in public conversations. We conduct a discourse analysis of 31 episodes from five popular data visualization podcasts. Podcasts are public-facing and inherently performative: guests manage impressions, articulate values, and model “good practice” for broad audiences. We use this performative setting to examine how legitimacy, identity, and practice are negotiated in community talk. We show that “inspiration talk” is operative rather than ornamental: speakers legitimize what counts, who counts, and how work proceeds. Our analysis surfaces four adjustable evaluation criteria by which inspiration is judged—novelty, authority, authenticity, and affect—and three operative metaphors that license different practices—spark, muscle, and resource bank. We argue that treating inspiration as a boundary object helps explain why these frames coexist across contexts. Findings provide a vocabulary for examining how inspiration is mobilized in visualization practice, with implications for evaluation, pedagogy, and the design of galleries and repositories that surface inspirational examples.
\end{abstract}




\maketitle
\raggedbottom

\section{Introduction}
Designers across domains routinely invoke inspiration to spark, shape, and justify ideas \cite{leifer_early_2014}. Data visualization design is no exception: practitioners draw on exemplars, prior projects, personal experiences, and media to generate and develop designs \cite{parsons_fixation_2021, bako_understanding_2022, bako_unveiling_2024, baigelenov_how_2025}. Inspiration has been studied in domains such as architecture \cite{cai_extended_2010, goldschmidt_creative_1998, mumcu_examining_2018}, interior design \cite{bettaieb_inspiration_2022}, knitwear design \cite{eckert_sources_2000}, interaction design \cite{chan_best_2018, halskov_kinds_2010}, and industrial design \cite{jagtap_inspiration_2017}, and only recently in visualization \cite{baigelenov_how_2025, bako_understanding_2022, bako_unveiling_2024, parsons_fixation_2021, yang_considering_2024}. Yet we still lack a discursive account of how inspiration is publicly constructed in the field---what counts as legitimate sources, how inspiration is narrated, and how such talk performs identity and boundary work.

This kind of account is important for several reasons. First, inspiration is widely recognized as central to the design process across a variety of design disciplines (e.g., \cite{eckert_sources_2000, goncalves_life_2021, mete_creative_2006}), and has more recently been foregrounded in visualization design as well (e.g., \cite{bako_understanding_2022, bako_unveiling_2024, baigelenov_how_2025}). While effective design does not require inspiration, both design researchers and practitioners consistently identify it as playing a significant role in creative work—from initial ideation to justifying design decisions. However, in both academic literature and practitioner discourse, inspiration often takes on different meanings in different contexts: it can refer to an external stimulus, a distinct design process stage, cognitive phenomena, affective experiences, or precedent knowledge. This multiplicity is not a mere theoretical confusion; it reflects the situated and dynamic ways in which inspiration actually functions in design practice. A discourse-analytic lens allows us to examine this multiplicity directly, revealing how practitioners actively construct and negotiate what inspiration means rather than treating it as a stable, pre-defined construct. 

Second, attending to how inspiration is talked about has practical implications for how visualization design is practiced, taught, and evaluated. Understanding how practitioners evaluate and frame inspiration can inform how critique is conducted, how precedents are introduced, and how design decisions are justified in teams and studios. It can also shape the design of repositories and example galleries so that they better support diverse creative practices rather than reinforcing a narrow notion of what “inspirational” work looks like. Making explicit the evaluative criteria and metaphors that circulate in inspiration talk can help students develop more flexible design identities and navigate critique, and it can help practitioners see how inspiration-related discourse functions not only as a description of the creative process but also as a means of constructing professional identity and negotiating community membership. 

Lastly, scholarship on data visualization design is relatively young, so its norms around inspiration, precedent, and originality are still actively being negotiated, unlike in more established fields (e.g., architectural design) with centuries of accumulated discourse. This makes it a particularly opportune moment to examine how inspiration is constructed in professional visualization design discourse.

We treat inspiration as both an external resource and a situated, discursive construct—something people actively build in and through conversations—since language is a vehicle for meaning-making, not only a window onto inner beliefs \cite{gee_introduction_2011}. A discourse-analytic lens foregrounds the rhetorical resources through which practitioners position inspiration and themselves—for example, metaphors, stance, positioning, narrative accounts, and legitimation strategies. This perspective complements prior studies (e.g., \cite{bako_understanding_2022, bako_unveiling_2024, baigelenov_how_2025}) by analyzing how practitioners talk about inspiration, rather than only which sources they use and why.

Public podcasts offer a tractable corpus for this inquiry. Episodes featuring highly visible practitioners (e.g., studio leads, book authors, and designers at prominent organizations) are designed for broad audiences and are inherently performative: guests manage impressions, articulate values, and tell polished stories. That performativity is analytically valuable for understanding how legitimacy, expertise, and “good practice” are constructed in community talk. In this study, we analyze 31 episodes from five popular data visualization podcasts recorded between 2019 and 2025, selected because they focus broadly on data visualization, remain active with fairly recent releases, and attract prominent visualization practitioners as guests.

Guided by this lens and corpus, we investigate the following research questions: RQ1: How do prominent visualization practitioners talk about inspiration in public podcast conversations? and RQ2: What discursive resources do they use to construct inspiration and professional identity? This work makes the following contributions: (1) A discourse-analytic account of inspiration talk in visualization practice, building on prior research on inspiration and visualization design practice; (2) an account of discursive resources—metaphors, positioning, narratives, and legitimation strategies used to construct inspiration; and (3) implications for how these constructions shape professional identities and community norms, with consequences for how inspiration is circulated and made meaningful within the field.

\section{Related Work} 
\subsection{Conceptualizing Inspiration}
Inspiration has a rich but fragmented theoretical history. It has been discussed in contexts ranging from Greek poetry and divine possession (e.g., \cite{murray_poetic_1981, goncalves_inspiration_2016}) to modern psychology (e.g., \cite{thrash_inspiration_2003}) and design theory (e.g., \cite{goncalves_inspiration_2016}). Despite this breadth, there is no widely agreed-upon, discipline-agnostic definition of inspiration. Early accounts emphasized divine or external origins, casting the person as a vessel for ideas \cite{murray_poetic_1981, goncalves_inspiration_2016}. More modern perspectives from psychology (e.g., \cite{thrash_creation_2021, thrash_psychology_2014, thrash_inspiration_2004}) reframe inspiration as a primarily cognitive internal phenomenon, but still emphasize that it must be \textit{evoked} by external stimuli \cite{thrash_inspiration_2003}. In design research, inspiration has largely been investigated as an external stimulus (e.g., design examples), while also being conceptualized as a \textit{process} through which stimuli are transformed into creative solutions \cite{eckert_algorithms_2000, goncalves_what_2014}. These models tend to present inspiration as a cyclical and iterative process (e.g., \cite{eckert_sources_2000, goncalves_inspiration_2016}).

Rather than proposing a single definitive account, we clarify the scope and boundaries of what we mean by inspiration. We treat inspiration broadly as anything that informs, stimulates, or shapes design decisions, including what might be referred to as influences, precedents, references, or exemplars. This expansive view is intentional: research on inspiration across disciplines (e.g., design research, psychology, HCI) reveals no settled consensus, and practitioners themselves use the term flexibly across contexts. For our purposes, inspiration can refer to external stimuli (e.g., design examples, nature, art), cognitive and affective states (e.g., sudden insights, excitement), processes (e.g., iterative exploration), and precedent knowledge (e.g., canonical work that sets community standards). We view these as not separate phenomena, but as different facets of how inspiration operates in design practice. We do \emph{not} consider as inspiration purely internal motivation affecting design decisions, post-hoc rationalization of completed work, and deliberate theft of other designers' work. 

\subsection{Inspiration in Visualization Design}
Besides discussions of inspiration-related tools and techniques (e.g., \cite{willett_perception_2022, owen_inspire_2023, judelman_aesthetics_2004, brehmer_generative_2022, he_vizitcards_2017}), few studies have embraced a practice-focused approach that investigates how inspiration ``works'' in visualization design. Bako and colleagues \cite{bako_understanding_2022} investigated what kinds of inspiration visualization designers look for and how they find it. They reported a wide variety of inspiration sources that designers draw on and techniques designers employ to incorporate those sources into their workflows. In their later work \cite{bako_unveiling_2024}, Bako and colleagues examined inspiration workflows in more depth, studying the influence of various factors on design solutions (e.g., when inspiration is given and the number and diversity of sources). 

While studying the processes and thinking of data visualization design practitioners, Parsons \cite{parsons_understanding_2022} described how designers use precedent knowledge as inspiration and discussed other sources from which designers draw. In later work, Parsons and colleagues \cite{parsons_fixation_2021} investigated a closely related concept to inspiration---design fixation---which is an unconscious tendency of designers to copy the elements of design examples they encounter in their eventual design solution. The study reported on various strategies that visualization designers employ to avoid fixation. Baigelenov and colleagues \cite{baigelenov_how_2025} examined visualization designers' subjective perceptions of inspiration and related phenomena, finding that designers often rely on diverse sources of inspiration beyond visualization examples. 

While these studies provide valuable accounts of practice, their data---mostly derived from interviews---were typically analyzed using thematic analysis, which treats designer-reported accounts of practice largely at face value. In itself, this is not a problem; however, if inspiration is a highly dynamic and situated construct, then it matters not only \emph{what} designers say but also \emph{how} they say it. Discourse analysis, which treats language as a form of social action, allows us to examine how inspiration is invoked, negotiated, and justified in talk, providing value beyond only face-value descriptions of inspiration and design process. Such a perspective can help in understanding how visualization practitioners themselves describe inspiration in terms of what metaphors, stances, narratives, and legitimation strategies they rely on to position inspiration in public conversations.

\subsection{Discourse Analysis in HCI}
Discourse analysis is ``the study of language-in-use'' \cite[p. 8]{gee_introduction_2011}, encompassing a variety of approaches that examine how language shapes identities, relationships, and norms, from fine-grained conversation analysis to studies of institutional power dynamics (e.g., \cite{foucault_orders_1971}). This family of methods is well-established in HCI scholarship, with researchers using it to investigate issues related to technology adoption (e.g., \cite{czech_independence_2023, nagele_subjectivities_2022}) and discussions around sensitive topics (e.g., \cite{kannabiran_how_2011, song_our_2024}), among others. In the visualization literature, however, discourse analysis has rarely been used, especially as a method of investigating design practice.

We draw on Gee's framework for discourse analysis \cite{gee_introduction_2011}, which treats language as a form of social action in which meaning is constructed through ``ways of saying, doing, and being'' \cite[p. 8]{gee_introduction_2011}. We do not adopt Gee's framework as a rigid coding scheme; rather, we use it as a sensitizing lens that informs our data analysis and provides vocabulary for describing how inspiration and professional identities are constructed in talk. We expand on our engagement with the framework in the Method section. Importantly, an interpretive method such as discourse analysis is well-suited to our data (see the next sub-section): podcast conversations are performative venues in which practitioners present themselves, articulate values, and position community norms, making them a rich site for examining how inspiration is invoked and negotiated.

\subsection{Podcasts as a Data Source}
Podcasts have increasingly become a medium for sharing and disseminating knowledge, and their contribution to knowledge production has been widely acknowledged. Across multiple fields including geography \cite{kinkaid_podcast-as-method_2021}, medicine \cite{zhang_how_2022, malecki_understanding_2019}, applied sciences \cite{welz_identifying_2020}, journalism \cite{apirakvanalee_telling_2023}, and education \cite{indahsari_using_2020}, to name a few, researchers have emphasized the role of podcasts in promoting learning and advancing disciplinary conversations. Podcasts are valuable because they shed light on current issues while bringing forward the perspectives of experts \cite{kulkov_leveraging_2024}. They also make data more tangible and emotionally resonant by conveying it through the tone and delivery of speakers’ voices \cite{kinkaid_podcast-as-method_2021}. Apirakvanalee and Zhai \cite{apirakvanalee_telling_2023} highlight how podcasts give voice to invited guests, amplifying diverse perspectives in the community. 

Recognizing this potential, scholars have proposed treating podcast ethnography as part of the broader ethnographic tradition, emphasizing that podcast data can be especially useful because of its flexibility across different contexts and timeframes \cite{lundstrom_podcast_2021, kinkaid_podcast-as-method_2021}. However, Kulkov and colleagues \cite{kulkov_leveraging_2024} caution that careful selection is necessary to ensure relevance and credibility.

Within the field of data visualization, podcasting and its audience have grown significantly, reflecting this broader trend. Against this backdrop, the discourse analysis of data visualization podcasts presents a valuable opportunity. These podcasts not only serve as vehicles for knowledge production but also allow us to examine how practitioners frame visualization practice and inspiration while addressing a broad audience. Analyzing them thus enables a richer understanding of how professional discourse in the visualization community is constructed, circulated, and made meaningful.

\section{Method}
We conducted a discourse analysis of episodes from popular visualization podcasts to understand how visualization practitioners construct and position inspiration in public discourse. In this section, we outline our approach and criteria for selecting the podcasts and specific episodes within those podcasts, and discuss our data analysis.  

\subsection{Podcast and Episodes Selection Criteria}
We selected five popular visualization podcasts for our analysis, namely (1) Data Viz Today by Alli Torban \cite{torban_data_2018}, (2) Explore Explain by Andy Kirk \cite{kirk_explore_2020}, (3) The Data Journalism Podcast by Alberto Cairo, Simon Rogers, and Scott Klein \cite{cairo_data_2021} (4) Data Stories by Enrico Bertini and Moritz Stefaner \cite{bertini_data_2012}, and (5) The PolicyViz Podcast by Jonathan Schwabish \cite{schwabish_policyviz_2015}. The podcasts were selected on the following criteria: (1) the podcast needed to focus broadly on data and information visualization (i.e., a general visualization podcast rather than one devoted to a narrow niche), (2) the podcast needed to be fairly active and have recent episodes, and (3) the podcast needed to be popular within the field and have prominent figures as guests (popularity and prominence of guests based on Google search, authors' own subjective opinions, and word of mouth). For the episodes, we focused only on episodes that retained the traditional podcast format of interview-like conversations between the host and guest(s); thus, we did not consider meta-episodes, recap and review episodes, promotional episodes, solo host episodes, and so on in our data collection and analysis. We also filtered out episodes where the guest was not primarily a visualization practitioner. Lastly, in the case of repeating guests across podcasts, we chose the episode for which we felt the data was richer for that particular guest and filtered them out from other episodes to avoid having the same practitioners appear multiple times in our data. 

\subsection{Data Source}
All of the episodes collectively featured 37 visualization practitioners, many of whom hold prominent roles in leading organizations such as major design firms and international news outlets. Given the nature of the data (i.e., podcast medium), these individuals are already naturally positioned as public voices of the community, and their professional contributions range from design work at globally recognized firms, visualization and graphics journalism at major newspapers, authorship of influential books on visualization design, and leadership of prominent visualization projects and design studios.

While analyzing podcasts is clearly engaging with secondary data, we believe that this data source presents a unique opportunity to investigate the discourse that these members of the visualization design community engage in, as they are typically very hard to recruit for a traditional interview-like study due to their high profile and busy schedules, yet at the same time they have high impact and reach within the community. 

\subsection{Data Analysis}
We stopped collecting data on July 1, 2025. Overall, we collected episode transcripts for 493 episodes across the five podcasts. The transcripts were obtained either through the podcast websites themselves or were generated using transcription services with manual correction based on the episode's audio or video. 

Podcasts represent a public medium; thus, no additional ethics approval was required. While podcast guests spoke publicly and are identifiable in source materials, we chose to anonymize individuals in our findings to foreground discursive patterns rather than individual speakers. Our focus is on how inspiration is collectively constructed and negotiated in public discourse, not on evaluating specific practitioners' views. Additionally, even though these conversations occurred in public venues designed to amplify practitioners' professional voices, guests did not explicitly consent to being research participants in an academic study (e.g., see \cite{dym_ethical_2020} for further discussion on the ethics of the analysis of public data). We wanted to avoid framing our analysis as evaluation or critique of individual practitioners' perspectives, which could create unintended consequences for their professional reputations. Still, as contextual information about speakers' professional backgrounds is analytically relevant for understanding these discourses, we provide detailed background information for the 13 practitioners whose quotes appear in our findings (see Table \ref{tab:demographics}). A complete list of all 31 analyzed episodes, including episode details, guest names, and direct links to source materials, is provided as supplementary material. This approach maintains our analytical focus on discourse patterns while ensuring transparency. Lastly, for readability purposes, we have lightly edited quotes by removing filler words without altering meaning.

\begin{table*}
  \caption{Details on podcast guests and their notable roles.}
  \label{tab:demographics}
  \centering
  \begin{tabular}{cccc}
    \toprule
    Practitioner & Work Context & Notable public role/achievements \\
    \midrule
    P3 & Energy sector & Local community leader  \\
    P4 & Private university & Professor, book author, speaker  \\
    P5 & Major science-related magazine & Book author, graphics editor  \\
    P7 & Design studio & Design studio founder and lead  \\
    P8 & Real estate company & Book author, speaker  \\
    P13 & Public university & Professor, award-winning designer   \\
    P16 & Major news organization & Award-winning designer   \\
    P19 & Major newspaper & Director   \\
    P29 & Tech company & CEO   \\
    P30 & Independent/Freelance & Creator of prominent visualization projects   \\
    P31 & Independent/Freelance & Award-winning designer, speaker, book author   \\
    P35 & Big news organization & Team co-leader   \\
    P37 & Major design consultancy & Award-winning designer, book author, speaker   \\

  \bottomrule
\end{tabular}
\end{table*}

To inform our analysis, we drew on Gee's \cite{gee_introduction_2011} framework for discourse analysis, which views language as a form of social practice. The main analytical tools of this framework are what Gee calls ``seven building tasks'', namely: significance, activities, identities, relationships, politics (the distribution of social goods), connections, and sign systems/knowledge. These tasks are then used to describe how language constructs meaning in situated contexts, for example by displaying how speakers position themselves and others, which moments they present as significant or insignificant, and how they frame concepts through their figured worlds (a shared mental model of how certain things are supposed to work).

While Gee's framework largely informed our analysis, we did not apply the framework very strictly and do not intend to present our work as a thorough application of this framework. Instead, we used this framework as a theoretical lens that guided us during our analysis and provided proper terminology for articulating the discursive patterns that we were observing in the data. For example, the concept of ``figured worlds'' helped us to describe how visualization practitioners framed inspiration in relation to their broader understanding of the field and community, while the notion of ``significance'' helped us in noticing instances of how visualization practitioners legitimized (or de-legitimized) inspiration-related moments. We believe that this flexible engagement with the framework allowed us to remain open to any emergent patterns in the data while still grounding our work in an established approach to discourse analysis.

We went through three rounds in our data analysis. In the first round, the first two authors collectively open-coded 15 episodes (3 latest episodes from each of the five podcasts), focusing on coding explicit and implicit references to inspiration, idea generation, and creativity broadly. We employed a strategy where we assigned very descriptive codes to our quotes. Although the podcast format involves back-and-forth conversations between the host(s) and guest(s), we were primarily interested in how the guests of those podcasts constructed and positioned inspiration in their discourse. Thus, while quotes from the hosts were often useful for understanding the context, our coding and analysis focused on what was said by the guests. To ensure consistency, the first two authors independently coded the same subset of the transcripts and then met continuously to discuss discrepancies, merging, or rephrasing codes as needed until consensus was reached. In this round we identified 205 codes. After the first round, all the authors had a discussion and collectively sorted the quotes into five discursive patterns: \textit{conceptualizing inspiration}, \textit{negotiating inspirational value}, \textit{positioning inspirational and creative self}, \textit{situated inspiration}, and \textit{constructing and circulating design knowledge}. In the second round, the first two authors coded an additional 10 episodes (next latest 2 episodes per podcast) with the new strategy. After the second round, all authors met for further discussion. We removed two of the discursive patterns (\textit{contextualized inspiration}, which was previously named \textit{situated inspiration}, and \textit{constructing and circulating design knowledge}), refined the naming of several discursive patterns and sub-patterns, and combined or shifted several sub-patterns. Overall, the remaining patterns (in their definition) stayed relatively the same, except for the \textit{conceptualizing inspiration pattern}, which we later reframed as \textit{mobilizing metaphors of inspiration}. At this point, we were reaching a point where no new significant discursive patterns were being identified; for thoroughness, we did a third round of analysis, where the first two authors coded an additional 6 episodes that were selected randomly to confirm saturation. In the end, we identified three discursive patterns and ten sub-patterns which we define and present in the next section.

\section{Findings}
Our analysis surfaced three recurrent ways practitioners construct inspiration in public talk: \textit{negotiating the value of inspiration}, \textit{positioning the creative self}, and \textit{mobilizing metaphors of inspiration}. Each pattern comprises several sub-patterns that function as discursive resources for legitimizing sources, claiming authority, and signaling norms.

\subsection{Negotiating the Value of Inspiration}
This pattern captures how practitioners evaluate when inspiration ``counts'', treating it as something to be justified rather than assumed. This evaluation takes multiple forms, which we present as sub-patterns.

\textbf{Legitimizing via Novelty.} One way practitioners negotiate the value of inspiration is through the degree of novelty it introduces to the field. In this perspective, inspiration is not just about producing high-quality or well-executed work, but also about pushing the boundaries and breaking away from established conventions. Practitioners frame inspiration as legitimate when it disrupts norms and signals novelty. For example, P31 reflected on evaluating entries for the Information is Beautiful Awards (a prominent source of inspiration in the community): \textit{``I was helping look through the shortlist for the 2022 Information is Beautiful Awards\ldots these are all really beautiful and well executed, but nothing is standing out\ldots we were trying to rank them, and\ldots from a visual perspective they were starting to feel very templated.''} 

Inspiration here is de-legitimized by its failure to disrupt formulaic community conventions, despite its solid technical execution. The evaluative language here contrasts ``beautiful'' and ``well executed'' with ``nothing is standing out'' and ``templated''. Specifically, the word ``templated'' positions repetition as a problem where inspiration loses legitimacy when it follows conventions too strictly. On the other hand, practitioners also acknowledge the limitations of such expectations and resist an absolutist discourse that demands constant novelty, as another practitioner (P37) pointed out: \textit{``not everything needs to be necessarily engaging to a point that you have never seen something like that before.''}

The phrase ``not everything needs to'' here sets a boundary and positions the speaker as a pragmatic designer in contrast to audience's implicit demands for constant novelty. The wording of ``never seen something like that before'' invokes an ideal of extreme novelty that is then downplayed as not necessary. This discursive move carves out a space for inspiration to also be recognized through craft and technical execution and not only through pushing the boundaries.

Together, these somewhat contradictory examples show how practitioners legitimize inspiration in relation to its novelty while negotiating its limits. In this sense, novelty and originality are central but not obligatory.

\textbf{Legitimizing via Authority.} Another way practitioners negotiate the value of inspiration is by tying it to recognized authority in the field. Inspiration is valued when it comes from credible, prominent, and/or well-known figures, teams, or institutions by borrowing legitimacy through association. For example, P19 described the stakes of producing work at a prominent organization: \textit{``You know that the president [of the United States] is going to probably read what you do. So you need to be at that level\ldots when I started my role at the Washington Post, it was like, okay, we are the innovation engine, we should be doing the cool things and we should be pushing the boundaries.''}

Here the speaker legitimizes the work (an implicit source of inspiration as a prominent organization in the field) that they do via its proximity to authority. The mention of ``the president'' discursively sets a bar for legitimacy where the imagined readership itself grants the weight to the work. Similarly, positioning the Washington Post as ``the innovation engine'' invokes a figured world where inspiration is validated via its association with a prestigious newsroom. On the other hand, practitioners borrow legitimacy by attaching themselves to prominent organizations whose authority is recognized in the community, even if they are not themselves formally part of those organizations. P13 mentioned the Pudding (a prominent digital publication) being an inspiration: \textit{``We are both huge fans of the Pudding and there was an obvious inspiration\ldots for the format of this piece\ldots we got a retweet at least, so I think that is a sign of approval.''}

\textbf{Validating Authenticity.} Practitioners also negotiate the value of inspiration by framing authenticity as a differentiating factor between legitimate inspiration and imitation. Inspiration is devalued when it involves mindless replication, but is validated when it leads to work that feels distinctively one's own. Discursively, this negotiation often involves drawing boundaries against practices of mindless and uncritical copying or overexposure to design examples. One practitioner (P3) described deliberately avoiding design examples: \textit{``Especially on LinkedIn\ldots when I used to participate in these challenges\ldots I would block myself from LinkedIn. I don't wanna see other people's work because we have so many people submitting. I don't wanna see other people's work because that would block my thinking.''}

The speaker positions exposure to too many design examples as threatening authentic creativity. The repeated mentions of ``I don't wanna see'' enacts a defensive stance in which the practitioner positions themself as guarding their originality against imitation. Here inspiration is de-legitimized if it comes through an unprotected mental space via absorption of other designers' work.

\textbf{Valuing via Affect.} Lastly, practitioners legitimize inspiration via affect where they value work that resonates with them emotionally or connects to personal experiences. Here inspiration is judged based on the feelings it evokes or embodies. One practitioner (P37) shared how their own experiences with COVID framed their perspective for a project:

\begin{quote}\textit{
    ``I have been suffering\ldots from long COVID\ldots and I have been battling with physical limitations, continuous flowing of symptoms of different kinds 24/7, pretty much for the past 4 years\ldots And, of course, I’ve collected a lot of data about it because this is what I do. And at some point, looking at the long COVID stories that were out there, I decided to try and publish something because the stories that I saw, and I think this is important in terms of what data can do, the story that I saw published were, you know, very moving, but at the same time, you would just read a blog post or an article and you’d read a list of 10 symptoms, 15 symptoms. And I think that for a healthy person, these might have felt like, well, okay, I’m tired as well after work\ldots But that is so far away from the experience of living 24/7 with a chronic illness.''}
\end{quote}

The practitioner here constructs a figured world in which their work (an implicit inspiration source) gains authority not from novelty or prestige but from lived experience. The immersive phrase of ``living 24/7 with a chronic illness'' positions the practitioner as both a patient and a designer which gives legitimacy to their work. Other practitioners (e.g., P5) highlight affective narratives in smaller scale aesthetic choices: \textit{``I love a good minimalist diagram as much as the next person, but sometimes they feel a little bit cold.''} 

The inspiration here is legitimized not by its technical execution, novelty, or its connection to a recognized authority but by its capacity to feel warm and inviting. The word ``cold'' here indicates that a design example lacking emotional resonance is devalued, regardless of its technical execution and stylistic polish. Collectively, these conversations illustrate how practitioners discursively position inspiration as valuable when it carries emotional weight.

\subsection{Positioning the Creative Self}
This pattern captures how practitioners discursively construct their creative identities in relation to inspiration by situating themselves through biography, expertise, or differentiation within their professional community. These constructions take multiple forms which we present as sub-patterns.

\textbf{Framing Inspiration via Expertise.} 
One way practitioners construct their creative identities is via positioning their unique professional expertise and backgrounds as distinctive framings that shape the way they approach inspiration and related practices. Here practitioners frame inspiration as shaped by their knowledge, training, background, and experiences, rather than as arbitrary. One practitioner (P35) describes the team environment in their organization which leads to a unique way of approaching inspiration: \textit{``Our process usually starts off with a lot of collaboration and creativity, so we try to organize regular brainstorm sessions\ldots weekly or biweekly\ldots we pitch ideas to each other, we share inspirations, we identify emerging trends and data sources. Those sessions really help to spark ideas. Each person in the team has their own individual background and preferences that they relate\ldots more to certain stories than other people, which results in a really great mix of different topics. Those sessions really spark ideas.''}

The practitioner here positions inspiration as a collaborative practice that is grounded in the team's collective expertise and distinct backgrounds. The emphasis on ``sharing inspirations'' and repeated phrase of ``really spark ideas'' foregrounds inspiration as a construct that is co-constructed collaboratively through the interplay of different identities rather than emerging from a single individual. The practitioner here highlights the differences between the members of the team which positions \textit{identity} as a resource that enriches inspiration. 

\textbf{Narrating Growth via Inspiration.}
Another way practitioners construct their creative identities is by positioning themselves as open, evolving, and authentic in their inspirational practices where they reveal their struggles, experimentation, and growth. Here practitioners construct a figured world in which their creative identity and expertise are not static, but are continuously reshaped through moments of inspiration. Inspiration becomes the medium through which the practitioners narrate their identity growth. For example, one practitioner (P16) described the surreal experience of shifting from being inspired by the work in the community to being considered an inspiration source themselves: \textit{``I mean\ldots having that validation of feeling respected by the DataVis community has been really nice\ldots someone on LinkedIn messaged to ask  how I did this, because they wanted to learn as well and that was wild to me because\ldots I am still in that position where\ldots some of the data visualizations I see and not understanding how they did it and would not even know how to start, so being on the opposite end of that was a little bit surreal.''} 

The practitioner here positions themselves as having two identities at the same time: the learner who is in awe of others' work and the recognized and established practitioner who inspires others. The phrases ``wild'' and ``surreal'' foreground inspiration as a marker of growth, signaling a shift in the practitioner's identity within the community. Another practitioner (P4) reflected on their stance towards a design piece they once criticized: \textit{``In the past, I harshly critiqued graphics and I should not have done that\ldots in the book\ldots I talk about a specific instance in which I reacted negatively to a graphic created by Amanda Cox from the New York Times as string graph. My first reaction to it was, this is terrible. This is really bad. But then in hindsight, I was completely wrong. This graphic is fantastic. It is really good. So, again the book is also a way for me to correct myself or correct past things that I do not feel proud of.''} 

Here the practitioner positions themself as an individual whose inspirational taste is evolving; they are able to re-frame their past judgments and re-situate their identity as a reflective and authentic practitioner within the community. Collectively, these conversations illustrate how practitioners use inspiration talk to show and narrate growth. 

\textbf{Signaling Belonging via Taste.} Lastly, practitioners construct their creative identities by displaying cultivated aesthetic preferences in their inspiration choices and aligning their inspirations with influential figures and recognizable styles. The practitioner's taste here becomes a discursive marker of community belonging and insider status. One practitioner (P29) explicitly connected their work to a prominent organization within the field (The New York Times): \textit{``For example, we just made a ``you draw it'' type template, similar to the ones that The New York Times pioneered.''} Here the practitioner aligns their work with a prominent institution, positioning community belonging as signaled through adopting recognizable design forms associated with respected outlets. Invoking The New York Times as having ``pioneered'' the technique presents a signal of legitimacy and insider status. Another practitioner (P31) reflected on their inspirational inclinations: \textit{``I realized that where I am most artistically inspired is in Japan and in East Asia.''} The practitioner here situates their inspirational tendencies within a specific culture, constructing a figured world in which belonging is expressed via affiliation with the aesthetic traditions of a particular culture. Together these examples illustrate how practitioners construct and position their creative identities within a particular community via taste. 

\subsection{Mobilizing Metaphors of Inspiration}
This pattern captures how practitioners frame and make sense of inspiration using metaphors and figurative language. Discursively these metaphors shape how inspiration is understood and what practices are legitimized (or de-legitimized) as a result. Different practitioners use different metaphors to understand inspiration---we present these framings as sub-patterns.

\textbf{The Spark.} One way practitioners make sense of inspiration is by framing it as a sudden ignition, spark, and/or ``aha moment''. This framing legitimizes serendipity and constructs a figured world in which creativity has some randomness to it. One practitioner (P30) describes inspiration as having an accidental and experimental nature: \textit{``A lot of inspiration just comes from accidents and playful explorations. That's the most important thing with just about any project, you have to explore and experiment.''} Here the practitioner foregrounds the accidental nature of inspiration by using the phrase ``the most important thing'' which legitimizes experimentation and treats creative space as serendipitous, requiring exploration via experimentation. Another practitioner (P8) reflects on their experience of coming up with creative ideas: \textit{``I usually find creative ideas when I am least expecting them, when I'm walking the dog or when I'm having a shower or when I'm on the treadmill\ldots I never think, right, okay I'm gonna walk the dog now and by the time I come back, I want an idea.''} Here the practitioner positions inspiration as something that is unpredictable and as something that often comes in spaces that are detached from work. The phrases ``I never think'' and ``when I am least expecting'' frames the inspiration as a spark that is serendipitous and outside of one's control. This framing positions inspiration as an outside force that implicitly legitimizes creative practices that do not directly relate to one's work (e.g., stepping outside of work space). Collectively, these examples illustrate how practitioners frame inspiration as a sudden spark which legitimizes serendipity in creative practices.

\textbf{The Muscle.} Another way practitioners make sense of inspiration is by framing it as a muscle---meaning, framing it as something that can be trained, built, and/or strengthened via daily routines and habits. This framing legitimizes ongoing collection, curation, and maintenance type of inspiration practices. For example, one practitioner (P7) reflected on what they think improves their creativity: \textit{``I also think that there's a daily practice\ldots that you can do to improve your creativity. I think the idea of researching inspiration, taking those notes, spending time, things that inspire you, you should do more of. So whether it's going to museum, whether it's watching movies, whether it's walking in nature, taking more showers. I don't know what inspires people, but whatever inspires them, I think we should do more of and collect those ideas somewhere\ldots And obviously the more you practice it\ldots the better you get.''} Here the practitioner constructs a figured world where inspiration can be trained (as a muscle) through discipline and routines. By using the words ``daily practice'' and ``do more of'', the practitioner frames creativity (and inspiration) less as a spark or serendipitous phenomenon, but rather as something that is more structured, deliberate,  and can be cultivated. 

\textbf{The Resource Bank.} Lastly, practitioners also make sense of inspiration by framing it as a resource bank which can be used for adapting, combining, and remixing other people's work without stigma or shame. This framing legitimizes re-using and re-combining existing ideas as authentic creative practices. For example, one practitioner (P8) directly stated that their creative ideas are adaptations of design examples that they have encountered: \textit{``I got my inspiration from someone else and that's very often the way it is\ldots if you can `steal like an artist'\ldots I get my creative ideas by adapting what I see\ldots I am gonna put my idea or my dataset with that person's idea and tweak it a little bit.''}  

Here the practitioner normalizes borrowing from others' work by dropping a famous reference ``steal like an artist'' which legitimizes reuse as an authentic creative practice. The practitioner here does not take a defensive stance, as in this framing of inspiration, reuse and remix are not only perfectly fine, but are actively legitimized and validated. Another practitioner (P35) further illustrates this stance: \textit{``We draw inspiration from so many amazing artists\ldots and we are like maybe we could draw inspiration from that and reshape it in a way that works for this topic..''}

The practitioner here legitimizes reuse and remix of inspiration by emphasizing the act of ``reshaping''. This framing positions remix as an authentic creative practice rather than simple copying. By highlighting how designers adapt prior ideas to fit a new context or topic, the practitioner frames borrowing from existing work as a natural and legitimate part of the creative process. Together, these accounts of design practice illustrate how practitioners construct a figured world in which borrowing from existing ideas is a perfectly legitimate approach to inspiration and creativity.

\section{Discussion}

Public “inspiration talk” is operative rather than ornamental: it performs \textit{legitimation} (what counts), \textit{identity work} (who counts), and \textit{practice-shaping} through metaphor (how work proceeds). We synthesize these results as \textit{adjustable evaluation criteria}—novelty, authority, authenticity, and affect—and as operative metaphors—\textit{spark}, \textit{muscle}, and \textit{resource bank}—that license different ways of working. We then consider implications for framing the construct, for creative identity and membership, for fixation and reuse, and for future work on how inspiration sediments into shared repositories, rubrics, and critique norms.

\subsection{Negotiated Framings of Inspiration}
Practitioners treat inspiration as a negotiated construct whose value is justified through \textit{adjustable evaluation criteria}: novelty, authority, authenticity, and affect. For some, inspiration is a random force that is outside of their control; for others, it is a controllable construct that can be strengthened with practice; for still others, it is a practical resource that can be reused, remixed, and adapted for one's purposes. These understandings differ, yet are still recognizable under the common label of inspiration. Rather than propose a new essence, we frame inspiration as a \textit{boundary object} \cite[p.~393]{star_institutional_1989}: plastic enough to fit local needs while retaining a recognizable identity across sites. Recent visualization research (e.g., \cite{otto_visualization_2024}) has similarly employed the boundary object lens to examine how visualization artifacts function across different stakeholder communities, suggesting the broader utility of this framing for understanding visualization practice. In our case, this view helps explain the lack of a single, domain-agnostic definition despite prior psychological accounts \cite{thrash_inspiration_2003,thrash_inspiration_2004} and process-oriented design framings \cite{eckert_adaptation_2003,goncalves_what_2014}. 

This multi-faceted, boundary object framing matters because it reorients how we study and support inspiration in visualization design and beyond. Rather than searching for a single definition or treating inspiration solely as an external stimulus (as in much of the prior work), it foregrounds inspiration as a construct that practitioners (and, to some extent, academics) actively construct and negotiate through discourse. This multiplicity is not a theoretical failure but a practical feature that allows inspiration to function across different contexts: individual ideation, collaborative critique, public presentation, and pedagogical scaffolding, while remaining recognizable. 

For researchers, this perspective opens several productive angles: examining how evaluation criteria shift across project phases; how metaphors license different practices; how identity work shapes whose inspiration counts; and how these constructions sediment into tools, repositories, and curricula. For practitioners and educators, recognizing inspiration's multi-faceted nature helps make its rhetorical work visible, enabling more strategic and reflective engagement with inspiration in creative practice. We expand on these practical implications in the following sections, examining consequences for creative identity and membership, fixation and reuse, and the circulation of design knowledge.

\subsection{Creative Identities and Inspiration}
Practitioners also use inspiration talk to do \textit{identity work}: claiming stances (expert/learner), narrating growth, and signaling membership through displays of taste (e.g., naming prestigious outlets or aligning with recognizable styles). In this sense, inspiration functions as a discursive resource for positioning oneself in the field, and not merely as a source of ideas. Prior work shows that designer identity is co-constructed with communities and contexts (e.g., \cite{gray_evolution_2014}); our data add that how one \emph{talks} about inspiration provides a compact vehicle for that co-construction. 

Two implications follow. First, heavy reliance on authority and taste as evaluation criteria risks reproducing exclusions; foregrounding authenticity and affect can expand what “counts” as legitimate inspiration. Second, making criteria explicit in critiques (e.g., “we are prioritizing novelty here”) can depersonalize disagreement and build students’ discursive flexibility. Investigating how the wider visualization community participates in this identity work—through awards, editorial policies, and public critique—remains an open direction (cf.\ \cite{kunrath_designers_2020,kunrath_social-_2019,tracey_reflection_2018,chivukula_identity_2021}).

\subsection{Inspiration Metaphors and Design Fixation}
Metaphors in practitioner discourse are not decorative; they are operational frames that license different activities \cite{lakoff_metaphors_1980}. Framing inspiration as a \textit{spark} legitimizes serendipity and wandering time; as a \textit{muscle}, disciplined routines and curated repertoires; as a \textit{resource bank}, adaptation and remix with provenance. These frames bear on long-standing concerns about design fixation \cite{jansson_design_1991,smith_fixation_1995,gero_fixation_2011,youmans_design_2014}. Attitudes toward imitation differ across design communities \cite{scolere_digital_2021,baigelenov_how_2025}; our data suggest a testable mechanism: (H1) communities that default to the \textit{resource bank} frame show more acceptance of adaptation and lower stigma around reuse; (H2) communities that foreground \textit{spark}/\textit{muscle} frames exhibit stricter anti-copying norms and different fixation profiles. These metaphorical repertoires may help explain why some design communities normalize remix while others stigmatize it. Comparative studies across domains (e.g., visualization vs.\ graphic design) could examine whether metaphor repertoires predict fixation attitudes and outcomes \cite{purcell_fixation_1993,purcell_design_1996}. Pedagogically, metaphor choice matters: a “spark-forward” curriculum may emphasize protecting fragile ideation from over-exposure, whereas a “bank-forward” curriculum can normalize adaptation with attribution and reflective justification.

\subsection{From Inspiration to Design Knowledge}
Our findings also point to inspiration as a precursor to \textit{circulating design knowledge}. Design researchers have long argued that design constitutes a ``third way of knowing'' distinct from science and the humanities \cite{nelson_design_2012}, which has led to attempts to theorize what counts as design knowledge. Practitioners archive iterations for later reuse, curate exemplar sets for students and clients, and learn from public critiques—practices that align with certain types of design knowledge, such as precedent and intermediate-level knowledge \cite{hook_strong_2012,Lowgren2013,stolterman_nature_2008,boling_nature_2021,nelson_design_2012}. We see an opportunity to trace \textit{micro-mechanisms of circulation} (newsletters, repositories, award rubrics, classrooms) and the roles that sustain them (custodian, curator, critic), and to analyze how evaluation criteria sediment into shared templates and norms. Clarifying where inspiration sits relative to precedent and other intermediate-level forms is a promising agenda rather than a settled claim. Is inspiration best understood as a type of precedent knowledge, a form of intermediate-level knowledge, or something distinct altogether? Future work could explore this positioning more systematically.

\subsection{Inspiration in Visualization Research}
Our account of adjustable evaluation criteria and operative metaphors complements recent visualization design studies that examine how designers seek, curate, and integrate design examples and inspiration into practice. Bako and colleagues' \cite{bako_understanding_2022} interview study surfaced the activities and challenges of design example search and use (criteria for usefulness; curation; fixation concerns), showing that designers balance effectiveness and aesthetics while selecting, merging, and modifying examples across projects. Our findings align with this balancing act but make visible two additional criteria that designers invoke in public talk—authority and authenticity—which do not figure as prominently in tool-centric accounts of example search. Making these criteria explicit helps explain why visually similar outcomes can be evaluated differently across venues or teams (e.g., prestige associations vs.\ protected voice).  

Bako and colleagues' later work \cite{bako_unveiling_2024} connects the timing and the type of design example exposure to what designers ultimately produce. When exposure was delayed, participants curated examples that were less topically similar and generated more varied designs. When design examples were schema-aligned, idea transfer into outcomes increased. Read through our lens, these effects could be understood as shifts in the evaluation criteria over time: early exploration often privileges novelty/variety (reaching wider), whereas schema-aligned reuse raises the salience of efficiency and feasibility, making resource-bank metaphors legitimate and idea transfer expectable rather than suspect. This re-framing connects outcome measures (variety, transfers) to the discursive justifications practitioners give in public.  

Closer to our topic, Baigelenov and colleagues \cite{baigelenov_how_2025} show that practitioners draw on diverse sources (examples, world phenomena, lived experience), use both active and passive inspiration practices, and are ambivalent about imitation—often accepting adaptive reuse when it carries attribution or identity. Our findings supply a mechanism: metaphors license practices (resource-bank legitimizes remix with provenance; spark supports protecting low-exposure ideation; muscle legitimizes cultivating disciplined repertoires), and evaluation criteria justify those choices (e.g., authenticity for lived-experience projects; authority for prestige-sensitive settings). Together, the studies suggest that inspiration is not just a stimulus but a field device for governing practice and membership.  

These works collectively point to precedent knowledge as a living substrate: Bako and colleagues' earlier work \cite{bako_understanding_2022} documents curation practices; Bako and colleagues' later work \cite{bako_unveiling_2024} quantifies idea transfer; our work via podcast conversation analysis reveals how designers publicly authorize precedent use (or police it) through criteria and metaphors. This triangulation supports our move to treat inspiration as a boundary object that travels between personal archives, community repositories, and award/review infrastructures—remaining “same enough” to recognize yet flexible enough to justify different uses across contexts.

Lastly, this study has focused on non-academic visualization practitioners. In practice, however, boundaries between academic and non-academic visualization work can be blurry, with individuals moving between roles or inhabiting hybrid positions. Future work could extend this discourse-analytic lens to academic contexts (e.g., conference presentations, design study publications, panel discussions) to examine how visualization researchers construct inspiration, and whether they emphasize different evaluation criteria or metaphors than those foregrounded in professional practice.

\subsection{A Discursive Angle for Studying Visualization Design Practice}
Beyond our specific results, focusing on practitioner \emph{discourse}—and, when appropriate, applying light-weight discourse-analytic techniques—usefully complements established visualization research approaches. A discursive lens makes tacit field rules visible: it reveals how designers publicly justify choices through adjustable evaluation criteria, how they negotiate membership through stance and displays of taste, and how operative metaphors license particular ways of working. It sits alongside existing methods by clarifying \emph{why} certain options become acceptable or obligatory before artifacts are finalized, and it resonates with interpretivist and critical/feminist turns in the community (e.g., \cite{akbaba_entanglements_2025}) by foregrounding power, voice, and standpoint.

For the broader community, crucially, this does not require a methodological overhaul. Visualization researchers can adopt small, tractable units of analysis within studies they already run—for example, noting which evaluation criteria (novelty, authority, authenticity, affect) are emphasized across phases; attending to how speakers position themselves and their imagined audiences; and periodically auditing which metaphor (spark, muscle, resource bank) is guiding activity and evaluation, and how shifting the frame would change decisions. The relevant traces are already abundant in our community: studio critiques and design-study check-ins, award jury statements, panel Q\&A, issue threads and pull requests, classroom discussions, and public talks and podcasts.

Finally, a discourse perspective opens cumulative research questions well suited to visualization research: How do criteria emphases vary across venues (journals, awards, classrooms)? Which metaphors dominate in subareas (e.g., narrative visualization vs.\ analytic dashboards), and with what effects on imitation stigma or fixation? Who gets to claim novelty or authority, and how do those claims shape review outcomes? Our goal is modest: not to teach a method, but to widen the imagination of what counts as rigorous visualization research on design practice. Selectively incorporating discourse analysis into existing visualization studies can surface mechanisms that are often reduced to “aesthetics”, and in doing so improve how we design, review, teach, and curate visualization work.

\subsection{Limitations}
As with any study, our approach has certain limitations. Because we did not generate these podcasts ourselves, our analysis relies on secondary data. In that sense, we had no control over what kinds of questions were asked in these conversations or which directions these conversations followed. The focus of each episode was inevitably shaped by the interests, interviewing style, and skills of the podcast hosts, which varied across podcasts. Formats of these podcasts also varied: some hosts focused more on guests' professional journeys, while others focused more narrowly on specific projects' details from a guest's repertoire. This led to some topics of potential interest to us (i.e., inspiration) not being covered. Our analysis was also very interpretive, as we were not the interviewers and could not ask clarifying questions. Furthermore, the role of the host also influenced what guests emphasized, both by the types of prompts provided and the conversational framing of inspiration-related topics.

Despite these constraints and limitations, we believe podcasts as a data venue offered several advantages that outweigh these limitations for our purposes. First, they provided access to high-profile practitioners who are otherwise difficult to recruit for traditional research interviews due to their visibility and busy schedules. Second, the performative, conversational, and publicly accessible nature of podcasts encouraged guests to reflect on and articulate their practices in polished ways, making their framing of inspiration-related issues analytically rich for discourse analysis. 

\section{Conclusion}
This paper has treated talk about inspiration in data visualization podcasts as data about how the field understands itself. Rather than assuming that inspiration is a private psychological state or a precondition for good design, we analyzed how prominent practitioners describe, justify, and contest inspiration in public, practice-focused conversations. In doing so, we showed that inspiration talk operates as a flexible discursive resource: it helps decide which sources and practices “count,” positions practitioners within overlapping communities, and orients work on concrete projects. Our analysis foregrounds the evaluative and metaphorical work that inspiration performs in data visualization design. Making these patterns explicit gives the community a language for examining its own norms—how examples are curated and rewarded, how critique is framed, how students are taught to think about precedent and reuse, and how tools and repositories implicitly steer attention toward particular kinds of “inspirational” work. Treating practitioners’ language about inspiration not as ornament, but as empirical material in its own right, opens up new ways of understanding---and ultimately reshaping---how visualization design is organized and justified in practice. 

\begin{acks}
This work was supported by NSF award \#2146228.
\end{acks}


\bibliographystyle{ACM-Reference-Format}
\bibliography{references}

\end{document}